\documentclass[twocolumn,pra,twocolumn,numerical,superscriptaddress]{revtex4-1}
\usepackage[a4paper]{geometry}
\usepackage{color}
\usepackage{bm}
\usepackage{amsmath}
\usepackage{amssymb}
\usepackage{graphicx}
\usepackage[unicode=true,pdfusetitle,
 bookmarks=true,bookmarksnumbered=false,bookmarksopen=false,
 breaklinks=false,pdfborder={0 0 1},backref=false,colorlinks=true]
 {hyperref}
\hypersetup{
 pdfborderstyle=,linkcolor=blue,urlcolor=blue,citecolor=blue}
\makeatletter
\usepackage{dcolumn}
\usepackage{bm}
\usepackage{epsfig}
\usepackage{braket}

\renewcommand{\fnum@figure}{FIG.~\thefigure}
\usepackage{upgreek}
\usepackage{placeins}
\usepackage{tikz}
\usepackage{pgfplots}
\pgfplotsset{compat=1.8}
\usetikzlibrary{external}
\date{\today}
\edef\flag{1}

\makeatother

\begin{document}
\title{Rotation of polarization of light propagating through \\
a gas of molecular super-rotors}

\author{Ilia Tutunnikov}
\affiliation{AMOS and Department of Chemical and Biological Physics, The Weizmann Institute of Science, Rehovot 7610001, Israel \looseness=-1}
\author{Uri Steinitz}
\affiliation{AMOS and Department of Chemical and Biological Physics, The Weizmann Institute of Science, Rehovot 7610001, Israel \looseness=-1}
\affiliation{Soreq Nuclear Research Centre, Yavne 8180000, Israel}
\author{Erez Gershnabel}
\affiliation{AMOS and Department of Chemical and Biological Physics, The Weizmann Institute of Science, Rehovot 7610001, Israel \looseness=-1}
\author{Jean-Michel Hartmann}
\affiliation{Laboratoire de M{\'e}t{\'e}orologie Dynamique/IPSL, CNRS, {\'E}cole Polytechnique, Institut Polytechnique de Paris, Sorbonne Universit{\'e},  {\'E}cole Normale Sup{\'e}rieure, PSL Research University, F-91120 Palaiseau, France \looseness=-1}
\author{Alexander A. Milner}
\affiliation{Department of Physics and Astronomy, The University of British Columbia, Vancouver V6T-1Z1, Canada}
\author{Valery Milner}
\email{vmilner@phas.ubc.ca}
\affiliation{Department of Physics and Astronomy, The University of British Columbia, Vancouver V6T-1Z1, Canada}
\author{Ilya Sh. Averbukh}
\email{ilya.averbukh@weizmann.ac.il}
\affiliation{AMOS and Department of Chemical and Biological Physics, The Weizmann Institute of Science, Rehovot 7610001, Israel \looseness=-1}

\begin{abstract}
We present a detailed theoretical and experimental study of the rotation of the plane of polarization of light traveling through a gas of fast-spinning molecules. This effect is similar to the polarization drag phenomenon predicted by Fermi a century ago and it is a mechanical analog of the Faraday effect. In our experiments, molecules were spun up by an optical centrifuge and brought to the super-rotor state that retains its rotation for a relatively long time. Polarizability properties of fast-rotating molecules were analyzed considering the rotational Doppler effect and Coriolis forces. We used molecular dynamics simulations to account for intermolecular collisions. We found, both experimentally and theoretically, a nontrivial nonmonotonic time dependence of the polarization rotation angle. This time dependence reflects transfer of the angular momentum from rotating molecules to the macroscopic gas flow, which may lead to the birth of gas vortices.  Moreover, we show that the long-term behavior of the polarization rotation is sensitive to the details of the intermolecular potential. Thus, the polarization drag effect appears as a novel diagnostic tool for the characterization of intermolecular interaction potentials and studies of collisional processes in gases.
\end{abstract}
\maketitle

\section{Introduction \label{sec:Introduction}}

Light traveling through matter is affected by the motion of the medium.
Already in 1818, A.-J. Fresnel considered this effect for a hypothetical ``aether
drag'' \citep{Fresnel1818}. The Fizeau experiment \citep{Fizeau1851}
seemingly demonstrated it; Fizeau showed that the phase and velocity of light change (being dragged) while propagating through a moving dielectric medium (flowing water) depending on the flow direction.
Later, J. J. Thomson argued that not only the phase (or velocity) but also the light's
polarization should be affected when propagating through a rotating aether
\citep{Thomson1885}, thus creating a transverse drag that leads
to the rotation of the plane of polarization. Eventually, the concept
of ``luminiferous aether'' was refuted by the Michelson-Morley experiment \citep{Michelson1887}.
E. Fermi considered the combination of Fizeau's
experiment and Thomson's ideas, and suggested that light traveling
through a rotating dielectric experiences a ``polarization drag'' \citep{Fermi1923}.
It took half a century to test the theory experimentally \citep{Jones1976}, 
and support it with further theoretical advances \citep{Player1976,Evans1992,Nienhuis1992}.
In the pioneering experiment \citep{Jones1976}, R. V. Jones  used a rapidly rotating glass cylinder that changed the laser's polarization direction by a few microradians. About a decade ago, the following experiment \citep{Franke-Arnold2011} enhanced the polarization drag by about four orders
of magnitude by using near-resonant slow light in a rotating
 ruby cylinder.

Detecting polarization rotation in gases is a much more difficult task,
primarily due to the medium's low density. 
J.-B. Biot measured the optical rotation caused by gases of chiral molecules \citep{Biot1819} under hazardous conditions, resulting in an accidental explosion of a 30 meter-long pipe filled with turpentine vapor.

Another major hurdle is the rapid rotation of a gaseous medium required to induce a measurable polarization drag.
However, recently we proposed a workaround \citep{Steinitz2020},
suggesting that instead of mechanically rotating a bulky dielectric
object as a whole, one could excite a fast unidirectional rotation
of individual microscopic particles. Current laser techniques (for recent reviews, see,
e.g., \citep{Ohshima2010,Fleischer2012,Lemeshko2013,Koch2019}) including
cross-polarized pulse pairs \citep{Fleischer2009,Kitano2009,Zhdanovich2011},
chiral pulse trains \citep{Zhdanovich2011}, polarization-shaped pulses
\citep{Karras2015},
and, especially, optical centrifuges \citep{Karczmarek1999,Villeneuve2000,Yuan2011,Korobenko2014,Owens2018,MacPhailBartley2020},
can bring molecules in the gas phase to very fast spinning.
When the molecules are excited to extremely high rotational states,
with their angular momentum reaching hundreds of $\hbar$, they become
super-rotors, which are rather resistant to collisions.
Gases of unidirectionally rotating super-rotors have been
studied both theoretically and experimentally.
They have many unique optical and kinetic properties; most relevant to this study are the rotational Doppler shift \citep{Korech2013,Korobenko2014,Steinitz2014} and inhibited rotational relaxation rate \citep{Korobenko2014,Khodorkovsky2015,Steinitz2016,Korobenko2018,Toro2013,Milner2014,Murray2015,Murray2016}.

Most recently, following the theoretical proposal \citep{Steinitz2020}, polarization drag (also known as the mechanical Faraday effect) in a molecular gas was experimentally demonstrated for the
first time \citep{Milner2021}. The experiments used the optical centrifuge pulses applied to various gases at ambient conditions. The molecules were spun up to high angular velocities, compensating for the gas's low density (compared to rotating solids). The results
demonstrated unprecedentedly high specific optical rotatory power \citep{CRC2021} (i.e., polarization rotation angle per unit propagation length per unit density), three orders of magnitude higher than
in the above record slow-light experiment \citep{Franke-Arnold2011},
and correspondingly almost nine orders of magnitude higher than
in Jones' work \citep{Jones1976}.

Here, we present a detailed theoretical analysis of the polarization
rotation of light propagating through a gas of fast-spinning molecules, and report new experimental results of time-resolved measurements of the polarization drag. 
Our study's central subject is the dynamics of the polarization rotation angle following the molecular excitation by the optical-centrifuge pulse. 
We argue that the measured time-dependence of the polarization drag reflects the exchange of angular momentum (due to intermolecular collisions) between the microscopic rotational degrees of freedom and the macroscopic gas flow. This time dependence provides indirect access to the details of generation of laser-induced gas vortices \citep{Steinitz2012,Steinitz2016}.     

The paper is organized as follows. In the next Sec. \ref{sec:Qualitative-2D}, we derive
an explicit expression for the polarization rotation angle using a
simplified two dimensional model, in which the molecules are confined
to a plane perpendicular to the light propagation direction. Section
\ref{sec:3D-case} contains practical formulas for the angle of polarization
rotation in the fully three dimensional case. Secion \ref{sec:Relaxation} focuses on the relaxation dynamics of the polarization rotation process. Section \ref{sec:Experimental} describes the experimental setup, presents the experimental results and compares them with the theoretical predictions.
Section \ref{sec:Conclusions} summarizes the paper.

\section{Qualitative description \label{sec:Qualitative-2D}}

Consider an electromagnetic wave propagating through a non-magnetic, homogeneous medium, whose macroscopic properties are time-independent. The wave equation for the electric field vector, $\mathbf{E}$ is:
\begin{equation}
\nabla^{2}\mathbf{E}-\frac{1}{c^{2}}\overset{\text{\text{\tiny\ensuremath{\bm{\leftrightarrow}}}}}{\boldsymbol{\varepsilon}}_{r}\frac{\partial^{2}\mathbf{E}}{\partial t^{2}}=0,\label{eq:wave-equation}
\end{equation}
where tensor $\overset{\text{\text{\tiny\ensuremath{\bm{\leftrightarrow}}}}}{\boldsymbol{\varepsilon}}_{r}$
is the relative permittivity, $c=1/\sqrt{\mu_{0}\varepsilon_{0}}$
is the speed of light in vacuum, $\mu_{0}$ is the magnetic permeability
of vacuum, and $\varepsilon_{0}$ is the permittivity of vacuum. In
the case of molecular gases, $\overset{\text{\text{\tiny\ensuremath{\bm{\leftrightarrow}}}}}{\boldsymbol{\varepsilon}}_{r}$
is related to the molecular polarizability tensor $\overset{\text{\text{\tiny\ensuremath{\bm{\leftrightarrow}}}}}{\boldsymbol{\alpha}}$ by
\begin{equation}
\overset{\text{\text{\tiny\ensuremath{\bm{\leftrightarrow}}}}}{\boldsymbol{\varepsilon}}_{r}=1+\frac{N}{\varepsilon_{0}}\braket{\overset{\text{\text{\tiny\ensuremath{\bm{\leftrightarrow}}}}}{\boldsymbol{\alpha}}},\label{eq:relative-permittivity}
\end{equation}
where $N$ is the number density, and $\braket{\overset{\text{\text{\tiny\ensuremath{\bm{\leftrightarrow}}}}}{\boldsymbol{\alpha}}}$
is the average (over all molecular orientations) polarizability. Substituting
Eq. \eqref{eq:relative-permittivity} into Eq. \eqref{eq:wave-equation}
yields 
\begin{equation}
\nabla^{2}\mathbf{E}-\frac{1}{c^{2}}\frac{\partial^{2}\mathbf{E}}{\partial t^{2}}=\frac{N\braket{\overset{\text{\text{\tiny\ensuremath{\bm{\leftrightarrow}}}}}{\boldsymbol{\alpha}}}}{\varepsilon_{0}c^{2}}\frac{\partial^{2}\mathbf{E}}{\partial t^{2}},\label{eq:wave-equation-final-1}
\end{equation}
Letting $\mathbf{E}=(E_{X},E_{Y})^T \exp[i(\omega t-kz)]$ (where superscript $T$ denotes transpose), allows us to rewrite
Eq. \eqref{eq:wave-equation-final-1} as 
\begin{equation}
\begin{split}\left(k^{2}-\frac{\omega^{2}}{c^{2}}\right)E_{X} & =\frac{N\omega^{2}}{\varepsilon_{0}c^{2}}[\braket{\alpha_{XX}^{(\omega)}}E_{X}+\braket{\alpha_{XY}^{(\omega)}}E_{Y}],\\
\left(k^{2}-\frac{\omega^{2}}{c^{2}}\right)E_{0Y} & =\frac{N\omega^{2}}{\varepsilon_{0}c^{2}}[\braket{\alpha_{YX}^{(\omega)}}E_{X}+\braket{\alpha_{YY}^{(\omega)}}E_{Y}],
\end{split}
\label{eq:coupled-wave}
\end{equation}
where $k$ is the wave number, the superscript $(\omega)$ means that
we consider the medium response at the same frequency as the driving
field. To simplify the solution of the system of equations in Eq.
\eqref{eq:coupled-wave}, we assume $\braket{\alpha_{XX}^{(\omega)}}=\braket{\alpha_{YY}^{(\omega)}}$ and
$\braket{\alpha_{YX}^{(\omega)}}=-\braket{\alpha_{XY}^{(\omega)}}$.
These assumptions will be justified in Sec. \ref{sec:3D-case}, when
considering the rotational excitation of molecules by the optical centrifuge.

Changing variables to $E^{\pm}=E_{X}\pm iE_{Y}$, we obtain the relative
permittivities for the two circular polatizations 
\begin{equation}
\begin{split}\varepsilon_{r}^{\pm}= &1+ \frac{N}{\varepsilon_{0}}[\braket{\alpha_{XX}^{(\omega)}}\mp i\braket{\alpha_{XY}^{(\omega)}}].\\
\end{split}
\label{eq:n-plus-n-minus}
\end{equation}
Finally, the angle of polarization rotation is given by \citep{Fowles1989}
\begin{align}
\Delta\Phi & =\frac{\omega L}{2c}\left(\sqrt{\varepsilon_{r}^{+}}-\sqrt{\varepsilon_{r}^{-}}\right)\approx-i\frac{\omega L}{2c}\frac{N}{\varepsilon_{0}}\braket{\alpha_{XY}^{(\omega)}},\label{eq:polarization-rotation}
\end{align}
where $L$ is the propagation length and the relative permittivities are very close to unity, as is typical in gases.

\begin{figure}[h]
\begin{centering}
\if\flag1\includegraphics{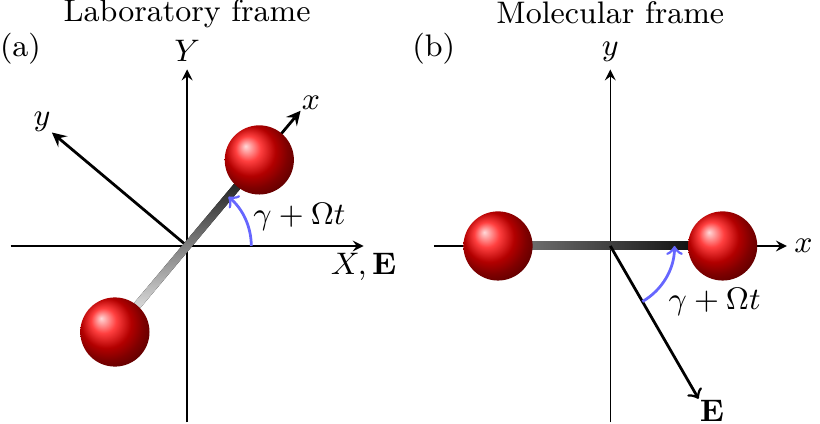}\else\include{FIG1}\fi
\par\end{centering}
\caption{Simplified 2D model. The molecule is restricted to rotate in the $XY$
plane. \label{fig:FIG1}}
\end{figure}

To obtain an explicit expression for $\braket{\alpha_{XY}^{(\omega)}}$ in the case of linear molecules,
we begin with a two-dimensional model in which the molecules rotate in
the $XY$ laboratory plane. We analyze the fully three dimensional
case, including intemolecular interactions, in the following sections.
Figure \ref{fig:FIG1}(a) shows a diatomic molecule rotating
in the $XY$ plane with an angular velocity $\Omega\mathbf{\hat{Z}}$, where
$\hat{\mathbf{Z}}$ is the unit vector along the $Z$ axis. At the input
plane, the electric field $\mathbf{E}$ of the probe light is modeled
by $\mathbf{\mathbf{E}}=E_{0}\exp(i\omega t)\hat{\mathbf{X}}$, where
$\hat{\mathbf{X}}$ is the unit vector along the $X$ axis. Prior
to probing, the molecules are assumed to be excited by another (pump)
laser pulse. Here, we neglect the effect of inhomogeneity stemming
from a non-uniform laser profile and the pulsed nature of the laser.
For the qualitative modeling, we assume that the molecule consists
of nuclei at a fixed relative distance and a single polarization charge
$q$ having a mass $m$. The nuclei attract the charge with a restoring electric
force which has components along the molecular
bond axis $x$ and along the perpendicular axis $y$
\begin{equation}
\begin{split}F_{x,\mathrm{r}} & =-k_{\parallel}x,\qquad F_{y,\mathrm{r}}=-k_{\perp}y\end{split}
.\label{eq:rest.-forces}
\end{equation}
This simple molecular model is capable of capturing the
essential physics, and agrees well with previous studies on the polarizability
of rotating objects \citep{Baranova1979,Pan2019}. For example, it demonstrates sensitivity to the anisotropy of the molecule.

In the rotating frame, there are inertial forces acting on the mass
\begin{equation}
F_{x,\mathrm{i}}=2m\Omega\dot{y}+m\Omega^{2}x,\;F_{y,\mathrm{i}}=-2m\Omega\dot{x}+m\Omega^{2}y,\label{eq:F-fict-2}
\end{equation}
including the Coriolis force (first terms) and centrifugal one (second terms). 
In the molecular frame [see Fig. \ref{fig:FIG1}(b)], the electric field components read
\begin{equation}
\negthickspace E_{x}=E_{0}e^{i\omega t}\cos(\gamma+\Omega t),\;E_{y}=-E_{0}e^{i\omega t}\sin(\gamma+\Omega t).\label{eq:Ex-Ey}
\end{equation}

\noindent Here, $\gamma+\Omega t$ is the instantaneous angle between the
$x$ axis and the electric field defined by
\begin{equation}
\cos(\gamma+\Omega t)=\hat{\mathbf{x}}\cdot\hat{\mathbf{E}}\;.\label{eq:cos(gamma)}
\end{equation}

Combining the restoring forces [see Eq. \eqref{eq:rest.-forces}],
inertial forces [see Eq. \eqref{eq:F-fict-2}], and the force due
to the electric field, yields the following coupled equations of motion
\begin{equation}
\begin{split}\negthickspace m\ddot{x}= & (m\Omega^{2}\!-\!k_{\parallel})x\!+\!2m\Omega\dot{y}\!+\!qE_{0}e^{i\omega t}\cos(\gamma\!+\!\Omega t),\\
\negthickspace m\ddot{y}= & (m\Omega^{2}\!-\!k_{\perp})y\!-\!2m\Omega\dot{x}\!-\!qE_{0}e^{i\omega t}\sin(\gamma\!+\!\Omega t).
\end{split}
\label{eq:App-eq.-of-motion}
\end{equation}
We solve this set of equations to get the driven steady-state solutions $x(t)$ and $y(t)$. By multiplying them by $q$, we obtain the induced dipole.
The dipole components in the laboratory frame
are found by projection on the $X$ and $Y$ axes.
Shortly after the end of the exciting pulses, the molecules become 
dispersed in angle due to different angular velocities and therefore 
we average the dipole moment over the angle $\gamma$.
The averaging nullifies the induced dipole
components oscillating at the shifted frequencies $\omega \pm 2\Omega$, which is consistent with the derivation of Eq. \eqref{eq:polarization-rotation}.
It is worth mentioning that in our experiments, special efforts were taken to eliminate the effect of residual linear birefringence (e.g., due to the quantum revivals) on the measured value of the polarization drag angle (see Sec. \ref{sec:Experimental} for details).

Expanding the dipole components (averaged over $\gamma$) in a power series up to the first order in $\Omega$ and substituting the polarizabilities of a non-rotating molecule
\begin{equation}
\begin{split}\alpha_{\parallel}(\omega) & =\frac{q^{2}}{k_{\parallel}-m\omega^{2}},\quad\alpha_{\perp}(\omega)=\frac{q^{2}}{k_{\perp}-m\omega^{2}},\end{split}
\label{eq:stationary-polarizabilities}
\end{equation}
we get
\begin{equation}
\begin{split}d_{X} & =E_{0}e^{i\omega t}\frac{\alpha_\parallel+\alpha_\perp}{2},\\
d_{Y} & =\frac{iE_{0}e^{i\omega t}m\omega}{q^2}(\alpha_{\parallel}-\alpha_{\perp})^{2}\Omega.
\end{split}
\label{eq:dipole-projections-X-Y}
\end{equation}
The off-diagonal component of the polarizability tensor is given by 
\begin{equation}
\braket{\alpha_{XY}}=\frac{\braket{d_{Y}}}{E_{0}e^{i\omega t}}=\frac{im\omega}{q^2}(\alpha_{\parallel}-\alpha_{\perp})^{2}\braket{\Omega},\label{eq:<alphaXY>}
\end{equation}
where the angle brackets denote averaging over the angular velocity, $\Omega$.
We can express the angle of polarization rotation using Eq. \eqref{eq:polarization-rotation} in terms of polarizabilities in Eq \eqref{eq:stationary-polarizabilities} and their derivatives (with respect to the driving frequency
$\omega$)
\begin{align}
\Delta\Phi & =\frac{1}{2}(\alpha_{\parallel}^{\prime}-\alpha_{\perp}^{\prime})\frac{\alpha_{\parallel}-\alpha_{\perp}}{\alpha_{\parallel}+\alpha_{\perp}}\frac{\omega NL}{2c\varepsilon_{0}}\braket{\Omega}.\label{DeltaPhi}
\end{align}
When $\alpha_{\parallel}=\alpha_{\perp}$, i.e. when the
tensor of polarizability is isotropic, $\Delta\Phi=0$. This is consistent
with the results of \citep{Baranova1979}. When $\alpha_{\perp}=0$,
i.e. the charge is restricted to move along the molecular axis, 
\begin{equation}
\Delta\Phi=\frac{\alpha_{\parallel}^{\prime}}{2}\frac{\omega N}{2\varepsilon_{0}}\frac{L}{c}\braket{\Omega},\label{DeltaPhi-2}
\end{equation}
which is the same as in \citep{Steinitz2020}. Equation \eqref{DeltaPhi}
shows that the polarization rotation angle is proportional to the
average molecular angular velocity, which is along the $Z$ axis in
the considered 2D case. Under certain conditions, the same holds for the three dimensional case too. 
Notice that we considered only the induced dipole oscillating at the frequency $\omega$ (the frequency of the input field), which is justified when the molecules are isotropically distributed in the $XY$ plane.

We can rewrite Eq. \eqref{DeltaPhi} in an
approximate form without using the polarizability components' derivatives $\alpha_{\parallel}^{\prime}$
and $\alpha_{\perp}^{\prime}$. We know from Eq.~\eqref{eq:stationary-polarizabilities} that $\alpha_{\parallel}^{\prime}{\alpha_{\parallel}}^{-2}=\alpha_\perp^\prime\alpha_\perp^{-2}$, and therefore   $\alpha_{\parallel}^{\prime}=f^{2}\alpha_{\perp}^{\prime}$,
where $f=\alpha_{\parallel}/\alpha_{\perp}$. Then, the derivative of the average 3-dimensional polarizability is $\bar{\alpha}^{\prime}=(2\alpha_{\perp}^{\prime}+\alpha_{\parallel}^{\prime})/3=\alpha_{\perp}^{\prime}(f^{2}+2)/3$.
Substitution of $\alpha_{\parallel}^{\prime}-\alpha_{\perp}^{\prime}=3\bar{\alpha}^{\prime}(f^{2}-1)/(f^{2}+2)$ in Eq. \eqref{DeltaPhi} leads to:
\begin{equation}
\Delta\Phi\approx\frac{3}{2}\frac{(f-1)^{2}}{f^{2}+2}\frac{\omega N\bar{\alpha}^{\prime}}{2\varepsilon_{0}}\frac{L}{c}\braket{\Omega}.\label{eq:App-Delta-Phi}
\end{equation}
The dispersion relation expressed by the difference of the group and phase refractive indices is $n_{g}-n_{\phi}=\omega N\bar{\alpha}^{\prime}/(2\varepsilon_{0})$,
 and therefore
%
\begin{align}
\Delta\Phi \approx \left[\frac{3}{2}\frac{(f-1)^{2}}{f^{2}+2}\right](n_{g}-n_{\phi})\frac{L}{c}\braket{\Omega}.\label{eq:App-Delta-Phi-2}
\end{align}

Using Eq. \eqref{eq:App-Delta-Phi-2}, we can estimate the specific rotatory power per unit average molecular angular velocity $[\alpha]/\braket{\Omega}=\Delta\Phi/(\rho L \braket{\Omega})$ (where $\rho$ is the mass density), for a variety of molecules. For example, for 
$\mathrm{O}_2$, $\mathrm{N}_2$, $\mathrm{CO}_2$ we get [in units of $\mathrm{deg\,(g/mL)^{-1}\,dm^{-1}\,(rad/ps)^{-1}}$]
$77$, $25$, $104$, respectively. To determine $f=\alpha_{\parallel}/\alpha_{\perp}$, the molecular data [($\alpha_{\parallel}-\alpha_{\perp}$) and $(\alpha_\parallel+2\alpha_\perp)/3$] was taken from \citep{Bridge1966}.  The dispersion relations were taken from \citep{Zhang2008,Kren2011} ($\mathrm{O}_2$), \citep{Borzsonyi2008} ($\mathrm{N}_2$) and \citep{Bideau-Mehu1973} ($\mathrm{CO}_2$). Notice that the measurements in \citep{Bridge1966} were done at a wavelength of $6328\,\textup{\AA}$, while in our experiments the wavelength of the probe light is $4000\,\textup{\AA}$. Nevertheless, since none of the example molecules has a resonance in the visible region of the spectrum, the obtained values of $f$ are good approximations.
At the average molecular angular velocity $\braket{\Omega}\sim 10\, \mathrm{rad/ps}$, achieved using typical optical centrifuge pulse (see Fig. \ref{fig:FIG3}), the specific rotatory power $[\alpha]$ may approach $1000\,\mathrm{deg\,(g/mL)^{-1}\,dm^{-1}}$, about three orders of magnitude higher than in \citep{Franke-Arnold2011}.

\section{Three dimensional case \label{sec:3D-case}}

We generalise  the analysis and find the
ensemble averaged polarizabilities when the molecules rotate in three
dimensions [the derivation is detailed in Appendix \ref{sec:Polarizabilities-3D-case},
see Eqs. \eqref{eq:App-alphaXX-1}-\eqref{eq:App-alphaYY}]:
\begin{align}
\braket{\alpha_{XX}} & =\frac{1}{2}[\alpha_{\parallel}(\omega)+\alpha_{\perp}(\omega)]\nonumber \\
 & -\frac{1}{2}[\alpha_{\parallel}(\omega)-\alpha_{\perp}(\omega)]\Braket{\frac{\Omega_{X}^{2}}{\Omega^{2}}},\label{eq:alphaXX-1}
\end{align}

\begin{align}
\braket{\alpha_{YY}} & =\frac{1}{2}[\alpha_{\parallel}(\omega)+\alpha_{\perp}(\omega)],\nonumber \\
 & -\frac{1}{2}[\alpha_{\parallel}(\omega)-\alpha_{\perp}(\omega)]\Braket{\frac{\Omega_{Y}^{2}}{\Omega^{2}}},\label{eq:alphaYY-1}
\end{align}
\begin{align}
\braket{\alpha_{XY}} & =-\frac{1}{2}(\alpha_{\parallel}-\alpha_{\perp})\Braket{\frac{\Omega_{X}\Omega_{Y}}{\Omega^{2}}}\nonumber \\
 & +\frac{i}{2}(\alpha_{\parallel}^{\prime}-\alpha_{\perp}^{\prime})\frac{\alpha_{\parallel}-\alpha_{\perp}}{\alpha_{\parallel}+\alpha_{\perp}}\braket{\Omega_{Z}},\label{eq:alphaXY-1}
\end{align}
where $\Omega_{X},\Omega_{Y},\Omega_{Z}$ are the components of the
molecular angular velocity, $\boldsymbol{\Omega}$, and $\Omega$
is its magnitude. Also, $\braket{\alpha_{XY}}=\braket{\alpha_{XY}^{\dagger}}$.
Equations \eqref{eq:alphaXX-1}-\eqref{eq:alphaXY-1} are valid when
the angular momentum of each molecule is conserved during
the measurement. In the pump-probe experiments in \citep{Milner2021}
and the present study, this criterion is satisfied, because the probe
duration is much shorter than the characteristic times of collisional relaxation (see Sections \ref{sec:Relaxation} and \ref{sec:Experimental}).

\begin{figure}[h]
\begin{centering}
\if\flag1\includegraphics{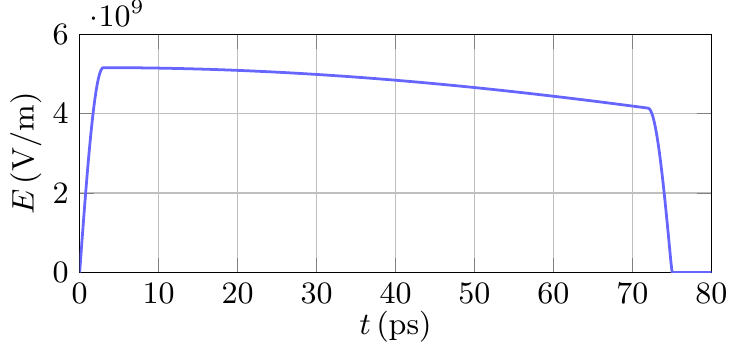}\else\include{FIG2}\fi
\par\end{centering}
\caption{Envelope of the optical centrifuge electric field. Here, the
duration of the optical centrifuge pulse is $\tau=75\,\mathrm{ps}$.
The peak amplitude corresponds to intensity of $I_{0\mathrm{c}}=3.5\,\mathrm{TW/cm^2}$.
\label{fig:FIG2}}
\end{figure}

To prepare an optically active molecular sample, i.e. a sample inducing
a measurable polarization rotation, we create molecular unidirectional
rotation in a gas of oxygen $(\mathrm{O}_{2})$ molecules. One of the most efficient
tools for this purpose is the optical centrifuge for molecules \citep{Karczmarek1999,Villeneuve2000,Yuan2011,Korobenko2014,MacPhailBartley2020} --  a laser pulse, whose linear polarization undergoes an accelerated rotation around its propagation direction. The electric field of optical centrifuge is modeled using
\begin{equation}
\mathbf{E}_{\mathrm{c}}=E_{0\mathrm{c}}f(t;\tau)\cos(\omega_{\mathrm{c}}t)[\cos(\beta t^{2})\hat{\mathbf{X}}+\sin(\beta t^{2})\hat{\mathbf{Y}}],\label{eq:optical-centrifuge-field}
\end{equation}
where $E_{0\mathrm{c}}$ is the electric field amplitude, $\beta$ is the angular acceleration of the centrifuge (here, $\beta=0.3\,\mathrm{rad/ps^2}$).
The function $f(t;\tau)$ defines the pulse envelope's time dependence (see Fig. \ref{fig:FIG2}), where $\tau$ is the duration of the pulse (here, $\tau=75\,\mathrm{ps}$). The electric field of the optical centrifuge pulse
polarizes the molecules. The induced dipole interacts with the same field resulting in a torque.
This torque drives the molecules to follow the vector $\mathbf{E}_{\mathrm{c}}$
that rotates in the $XY$ plane. Depending on their orientation and
angular momentum at the beginning of the laser pulse, a fraction of
the molecules is ``captured'' by the optical centrifuge and those
are accelerated to higher angular velocities. The axes of these molecules tend to be perpendicular to the $Z$ direction, so they behave more as in the two-dimensional case.

To simulate the rotational-translational dynamics of the molecules
at ambient conditions we use classical molecular-dynamics simulations (CMDS), which are carried out as described in \citep{Hartmann2012,Steinitz2016} where more
information can be found. Briefly, one first initializes the center of
mass positions (randomly, but not too close to each other), and the
velocities as well as the molecular axis orientation and angular momentum
vector so that they verify the Boltzmann statistics for a gas at thermodynamic
equilibrium. These four quantities define the state of each molecule, and
are then propagated in time as follows. First, at each time step,
we calculate the force and torque applied to each particle by its
neighbors. This is done using the known positions and orientations
of all molecules and an input intermolecular potential (from \citep{Bouanich1992},
for $\mathrm{O_{2}}$--$\mathrm{O_{2}}$). During the centrifugation
process, for each particle we compute the torque exerted by
the laser's electric field according to the (known) instantaneous
orientation of the molecular axis and the polarizability anisotropy $(\alpha_{\parallel}-\alpha_{\perp})$. We add it to the torque exerted by the neighboring molecules. After computing the forces and torques, the molecules positions, orientations, linear and angular velocities are evolved classically through the time step. The
CMDS simulations, which use periodic boundary conditions, nearest neighbors spheres
and the Verlet algorithm \citep{Allen1987}, provide a full description
of the translational and rotational states of each molecule at all
times. From these, distributions and mean values of various dynamical
quantities, including the averages appearing in Eqs. \eqref{eq:alphaXX-1}-\eqref{eq:alphaXY-1},
can be derived.

Our CMDS simulations show that at the end of the  optical centrifuge
excitation, $\braket{\Omega_{X}^{2}/\Omega^{2}}\approx\braket{\Omega_{Y}^{2}/\Omega^{2}}$
and $\braket{\Omega_{XY}^{2}/\Omega^{2}}\ll1$. Accordingly, $\braket{\alpha_{XX}}\approx\braket{\alpha_{YY}}$,
and $\braket{\alpha_{XY}}\approx-\braket{\alpha_{XY}}$. This verifies the assumptions we made in the derivation of the formula for the polarization
rotation angle in Eq. \eqref{eq:polarization-rotation}.
Therefore, we may substitute the formula in Eq. \eqref{eq:polarization-rotation}
with $\braket{\alpha_{XY}}$ from Eq. \eqref{eq:alphaXY-1} to obtain 
\begin{equation}
\Delta\Phi\approx\frac{1}{2}(\alpha_{\parallel}^{\prime}-\alpha_{\perp}^{\prime})
\frac{\alpha_{\parallel}-\alpha_{\perp}}{\alpha_{\parallel}+\alpha_{\perp}}\frac{\omega NL}{2c\varepsilon_{0}}\braket{\Omega_{Z}},\label{eq:3D-polarization-rotation}
\end{equation}%
which is similar to the expression obtained in the 2D case {[}see Sec.
\ref{sec:Qualitative-2D}, Eq. \eqref{DeltaPhi}{]}.
Alternatively, the polarization rotation angle may be written as
\begin{equation}
\Delta\Phi\approx\left[\frac{3}{2}\frac{(f-1)^{2}}{f^{2}+2}\right](n_{g}-n_{\phi})\frac{L}{c}\braket{\Omega_{Z}},
\end{equation}%
similarly to Eq. \eqref{eq:App-Delta-Phi-2}.

\section{Relaxation dynamics \label{sec:Relaxation}}
\begin{figure}
\begin{centering}
\if\flag1\includegraphics{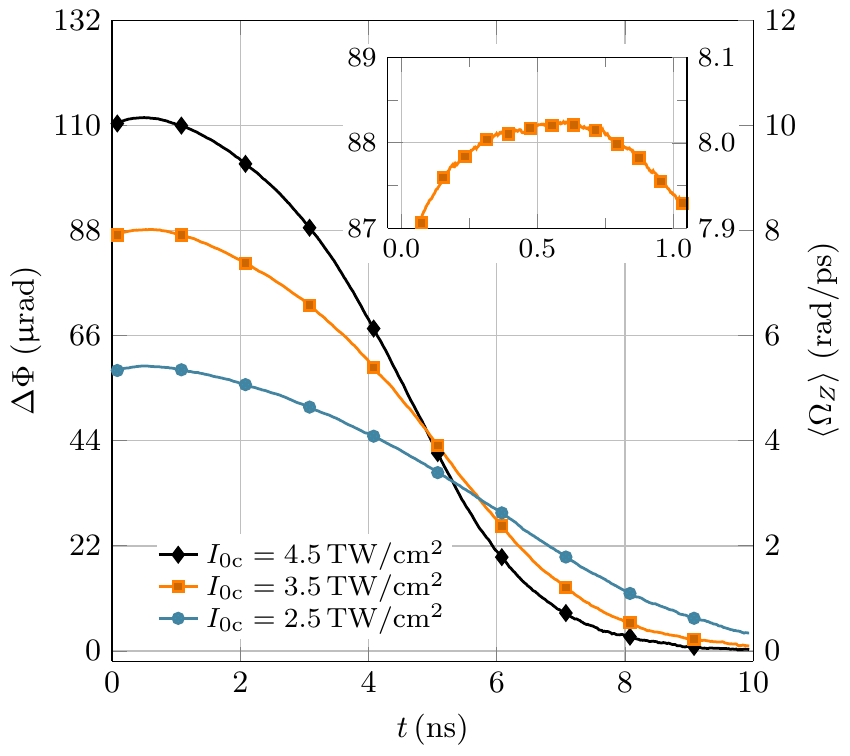}\else\include{FIG3}\fi
\par\end{centering}
\vspace{-3mm}
\caption{Average $Z$ projection of the molecular angular velocity (right ordinate)
and angle of polarization rotation (left ordinate) after excitation by the optical centrifuge pulse calculated for a gas of $\mathrm{O}_{2}$
molecules. Pulse
duration is $\tau=75\,\mathrm{ps}$, centrifuge angular acceleration is $\beta=0.3\,\mathrm{rad/ps^2}$.
Initial temperature is $T=300\,\mathrm{K}$, pressure is $0.5\,\mathrm{bar}$.
Propagation length is $L=1\,\mathrm{mm}$. \label{fig:FIG3}}
\end{figure}

Figure \ref{fig:FIG3} shows the computed average projection of the molecular
angular velocity on the $Z$ axis, $\braket{\Omega_{Z}}$ (right ordinate) following the optical centrifuge pule excitation for various peak intensities. The angle of polarization rotation,
$\Delta\Phi$ (left ordinate) was obtained using Eq. \eqref{eq:3D-polarization-rotation}
with frequency-dependent polarizabilities taken from \citep{Hettema1994} (for
details, see Appendix \ref{sec:Appendix-Frequency-dep-pol}). 
After the pulse, the total angular momentum
(as a vector) of all the molecules is conserved. It is the sum of rotational angular momenta of each molecule about its center of mass and angular momenta of the centers of mass about the origin (center of the
optical centrifuge focal plane).
The time-dependence of $\braket{\Omega_{Z}}$ (or, equivalently, of the average
$Z$ component of the angular momentum, $\braket{L_{Z}}$) seen in
Fig. \ref{fig:FIG3} is the indirect evidence of the angular momentum transfer
process. Due to intermolecular collisions, the rotational angular
momentum is transferred into the center of mass angular momentum.
Under certain conditions, this process leads to the birth of a macroscopic
vortex, which was theoretically analyzed in \citep{Steinitz2012,Steinitz2016}.
\begin{figure}
\begin{centering}
\if\flag1\includegraphics{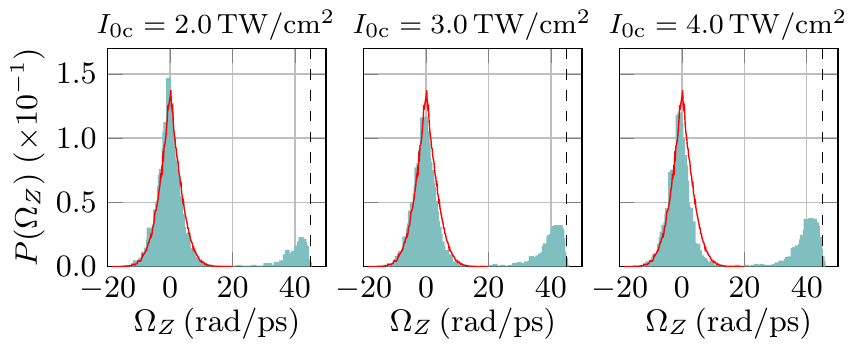}\else\include{FIG4}\fi
\par\end{centering}
\vspace{-3mm}
\caption{Distributions of $Z$ projection of the molecular angular velocity at the end of the optical centrifuge
pulse. Pulse parameters are the same as in Fig. \ref{fig:FIG3}. The expected
terminal angular velocity (denoted by vertical dashed line) is $[\Omega_{Z}]_{\mathrm{exp}}\approx2\beta\tau=45\,\mathrm{rad/ps}$.
Initial thermal distribution is denoted in red. \label{fig:FIG4}}
\end{figure}%

\begin{figure}[h]
\begin{centering}
\if\flag1\includegraphics{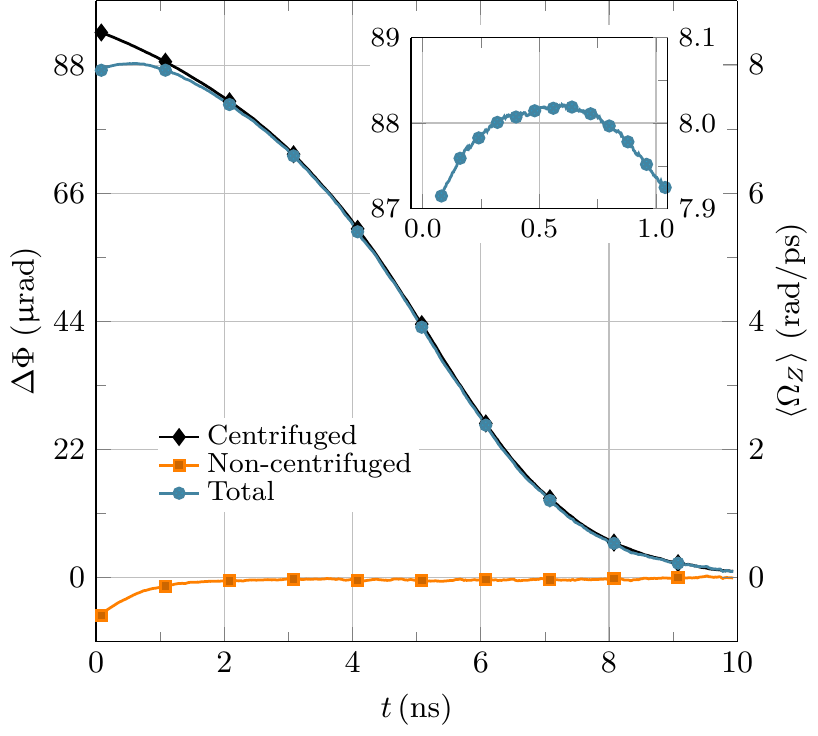}\else\include{FIG5}\fi
\par\end{centering}
\vspace{-3mm}
\caption{Average $Z$ projection of the molecular angular velocity (right ordinate)
and angle of polarization rotation (left ordinate) for centrifuged and non-centrifuged molecules (see the text). Here, $I_{0\mathrm{c}}=3.5\,\mathrm{TW/cm^2}$. The rest of the parameters are the same as in Fig. \ref{fig:FIG3}.
\label{fig:FIG5}}
\end{figure}

Notice that for $t<1\,\mathrm{ns}$ there is a visible nonmonotonic behavior (a ``bump'') in the
graphs of $\braket{\Omega_{Z}}$ and $\Delta\Phi$ (see the inset in
Fig. \ref{fig:FIG3}). To explain why they both first increase before thermalization
and decay to zero, let us consider the results shown in Fig.
\ref{fig:FIG4}. The latter displays the distributions of $\Omega_{Z}$ at the end of the
centrifuge pulse for
various peak intensities (calculated under collision-free conditions).
We can identify the following main features, confirmed by multiple experimental observations (e.g. \citep{Korobenko2014,MacPhailBartley2020}). (i) The distribution is bimodal
with a clear separation between the molecules that have been caught
by the centrifuge (right hand side peak) and those which have not
(left hand side peak). (ii) The distribution of the accelerated molecules
peaks, as expected, near the terminal value of the centrifuge angular
speed ($2\beta\tau$). (iii) As the energy and peak intensity of the
pulse increase, so does the relative number of centrifuged molecules
(at the fixed initial temperature used here, $T=300\,\mathrm{K}$).
This comes from the fact that the centrifuge becomes more and more
efficient in capturing (and keeping) molecules thanks to the laser-induced torque increasing with the amplitude
of the laser electric field. This explains the increase of $\Omega_{Z}$
(and thus $\Delta\Phi$) with the pulse intensity shown in Fig. \ref{fig:FIG3}.
For a detailed theoretical analysis of centrifuge-driven molecular
dynamics, see \citep{Steinitz2016,Armon2016,Armon2017}. (iv) 
Last but not least, comparing the distribution of the non-centrifuged molecules with the initial Boltzmann distribution (shown in red), we find that the centrifuge preferentially catches those molecules that were initially rotating with the polarization vector (i.e., have a positive $\Omega_{Z}$).
This implies that, when the
centrifuge is turned off, $\braket{\Omega_{Z}}$ for the non-centrifuged
molecules (integration over the left hand side peak) is negative,
while $\braket{\Omega_{Z}}$ for the centrifuged molecules (integration
over the right hand side peak) is positive. This last feature,
together with the fact (discussed below) that the collisional thermalization
(i.e. isotropization) of the rotational angular momenta is faster
for slowly spinning molecules than for very fast rotors, explains
the presence of a bump in the drag angle at early times shown in the
insert of Fig. \ref{fig:FIG3}. Figure \ref{fig:FIG5} separately shows $\braket{\Omega_{Z}}$
for centrifuged and non-centrifuged molecules. 
The molecules were divided into the two groups based on their
angular velocity at the end of the laser pulse, $\Omega_{Z}>20\,\mathrm{rad/ps}$
(for centrifuged) and $\Omega_{Z}<20\,\mathrm{rad/ps}$ (for non-centrifuged), in agreement with Fig. \ref{fig:FIG4}. After the excitation, $\braket{\Omega_{Z}}$ for the non-centrifuged
is negative, while
the value for the accelerated molecules the value is positive. 
Figure \ref{fig:FIG5} also shows that the decays rates are very
different for the two groups of molecules, significantly smaller for the centrifuged molecules than for
the non-centrifuged ones. This results in a ``bump'' in the plot of total $\braket{\Omega_{Z}}$ obtained by summing the two contributions. The quicker thermalization of the slowly
spining molecules compared to the fast ones, is due to the
fact that intermolecular interactions are more efficient in changing
the angular momentum of a slow rotor than when the molecule is spinning
very fast (a \textquotedblleft super-rotor\textquotedblright{} effect
discussed in \citep{Korobenko2014,Khodorkovsky2015,Steinitz2016}). Indeed, in this
latter case, only very strong (and thus rare) intermolecular collisions
involving a high relative speed and a small impact parameter are efficient. For completeness, we note that similar
arguments were invoked in the study of the polarization of the electronic
spin of centrifuged $\mathrm{O}_{2}$ molecules \citep{Floss2018}
in order to explain the nonmonotonic time dependence of the collisional dissipation
of macroscopic magnetization.
\vspace{-5mm}
\begin{figure*}
\begin{centering}
\includegraphics[width=0.85\textwidth]{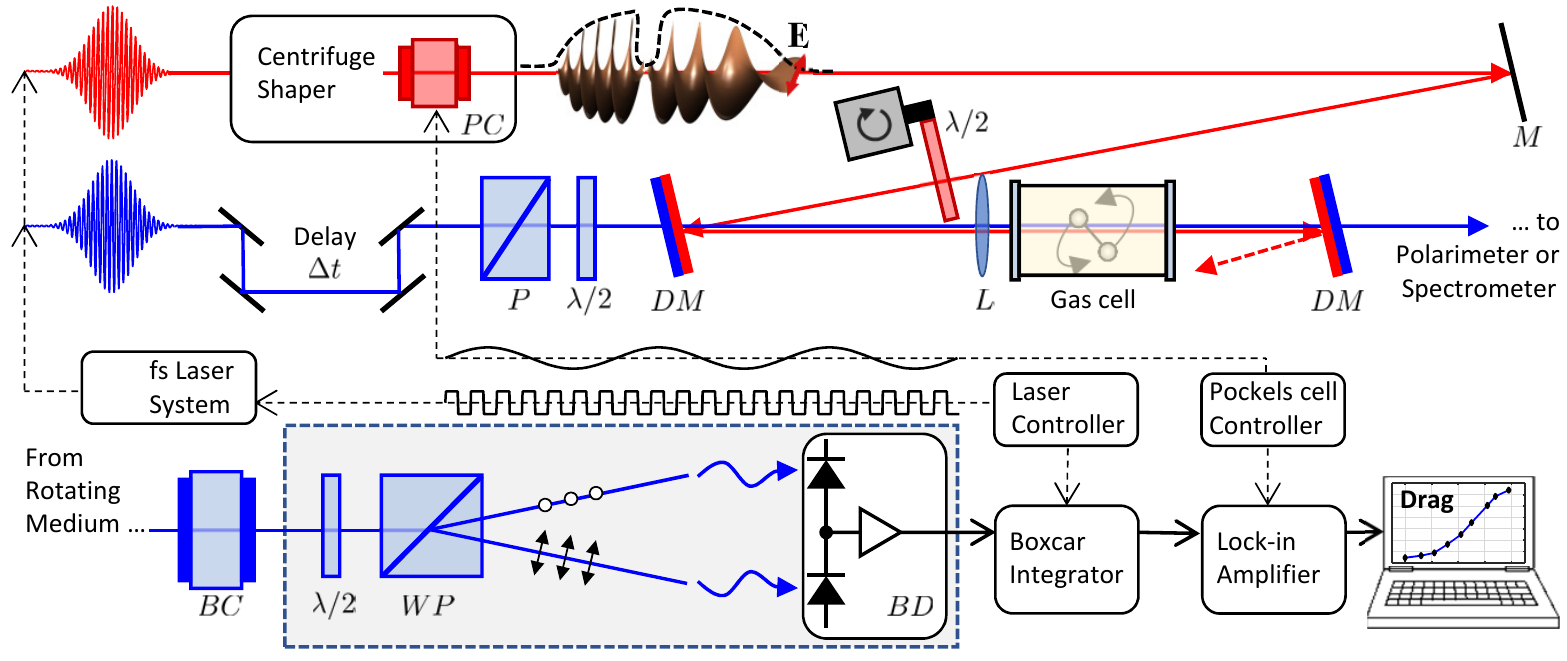}
\par\end{centering}
\caption{Scheme of the experimental setup. Top: femtosecond pulses with the
central wavelength of $792\,\mathrm{nm}$ (upper, red) and $398\,\mathrm{nm}$
(lower, blue) are used for creating the centrifuge and the probe pulses,
respectively. The pulses are shaped, delayed with respect to one another,
combined in a collinear geometry and focused in a gas cell. Bottom:
after passing through the gas sample, probe pulses are filtered out
from the centrifuge light and sent to the time-gated polarization
analyzer, implemented with a boxcar integrator. $PC$: Pockels cell,
$\lambda/2$: zero-order half-wave plate, $P$: polarizer, $M$: metallic
mirror, $DM$: dichroic dielectric mirror, $L$: lens, $BC$: Berek
compensator (Newport 5540), $WP$: Wollaston prism (Thorlabs WP10-A),
$BD$: balanced detector (Thorlabs PDB220A2). Alternatively, the probe
pulses may be sent to a Raman spectrometer to characterize the rotation
of the centrifuged molecules. \label{fig:FIG6}}
\end{figure*}
\section{Experimental demonstration \label{sec:Experimental}}
We have developed a sensitive experimental setup for the polarization
drag measurements. It is schematically shown in Fig.~\ref{fig:FIG6}
and has been described in detail in a recent publication \citep{Milner2021}.
It consists of an optical centrifuge,
an oxygen gas cell, and the detection optics. To achieve the required
sensitivity of $\sim1\,\upmu\mathrm{rad}$, the conventional centrifuge
was modified in the following way.

A stationary quarter-wave plate at the output of the centrifuge pulse
shaper was replaced by a Pockels cell ($PC$), which enabled us to
modulate the direction of the centrifuge rotation between CW and CCW
at the frequency of $37\,\mathrm{Hz}$. A lock-in amplifier was employed
to record the signal at the modulation frequency. All mirrors delivering
the centrifuge pulses to the gas sample ($M,\;DM$) were oriented as
close to the normal incidence as possible (unlike the traditional
$45^{\circ}$ geometry). This minimized the distortions of the centrifuge
polarization, which introduced a preferential axis of the field polarization
and produced an undesired linear anisotropy in the sample. The latter
resulted in the unpredictable rotation of the probe polarization,
which contaminated (and often dominated) the polarization drag signal.
To further suppress any residual linear anisotropy (e.g. due to the
sudden rising edge of the centrifuge), we passed the centrifuge beam
through a half-wave plate, mechanically rotated at a few revolutions
per second. In addition to flipping the centrifuge direction, which
was taken into account in our analysis, the rotating wave plate randomized
the direction of the rising edge, averaging its effect on the polarization
rotation to zero. The successful cancelation of these linear artifacts
was ensured by the appropriately long time constant ($\sim1\,\mathrm{s}$)
of the lock-in amplifier.

To eliminate additional systematic errors, the method of controlling
the rotational frequency of the centrifuged molecules has also been
modified. In all of our previous studies, this frequency of the super-rotors
was controlled by truncating the centrifuge pulse in time, which resulted
in an early termination of the accelerated molecular rotation and
correspondingly to a lower rotational frequency. This method involved significant
changes of the total pulse energy by up to $50\%$. Given the high
centrifuge intensities, large variations in the pulse energy can lead
to changes in the local heating of optical elements, potentially affecting
their birefringence and modifying the polarization of the probe pulses
passing through them. To suppress any effects of the pulse energy
on the sensitive polarimetry setup, we developed a new technique of
centrifuge ``piercing'', described in detail in \citep{Amani2021}.
As schematically illustrated in Fig.~\ref{fig:FIG6}, a short $2\,\mathrm{ps}$
notch was introduced in the field envelope of an optical centrifuge
by means of a spectral filter in the centrifuge pulse shaper. The
notch interrupts the accelerated rotation of molecules at any desired
rotational frequency. Controlling that frequency was executed by moving
the notch position in time, which was accompanied by less than 2\%
variation in pulse energy.

The centrifuge pulses were focused in a cell filled with oxygen gas
at room temperature and at pressure of $0.5\,\mathrm{bar}$. The focusing
lens $L$ with a focal length of $10\,\mathrm{cm}$ provided a length
of the centrifuged region of about $1\,\mathrm{mm}$ and a peak intensity
of up to $5\,\mathrm{TW/cm^{2}}$. Probing the gas of
oxygen super-rotors was done with short probe pulses (pulse lengths
of $3\,\mathrm{ps}$) delayed with respect to the centrifuge. The
probe pulses were derived from the same laser system, spectrally narrowed
down to the bandwidth of $0.1\,\mathrm{nm}$, and frequency doubled
to $400\,\mathrm{nm}$ for the ease of separating them from the excitation
light. Care was taken to make the probe focal spot smaller than the
corresponding size of the centrifuge beam to minimize the effects
of pointing instability.

The detection scheme consisted of two alternate channels: the Raman
channel and the polarization drag channel. In the former, the probe
pulses (filtered out from the centrifuge light) were circularly polarized
and sent to a spectrometer. Coherent scattering from the centrifuged
oxygen molecules resulted in Raman spectra with well-resolved peaks,
corresponding to individual rotational quantum states \citep{Korobenko2014}.
The magnitude of the Raman shift was translated to the rotational
frequency, which was set to the desired final value by moving the
position of the frequency piercing in the centrifuge shaper \citep{Amani2021}.

In the second detection channel, a very sensitive method for measuring
the small degree of polarization rotation was implemented by means
of an optical configuration depicted inside the dashed gray rectangle
at the bottom of Fig. \ref{fig:FIG6}. A Berek compensator ($BC$)
and a half-wave plate were used to align the (in this case, linear)
probe polarization at $45^{\circ}$ with respect to the axes of a
Wollaston prism. This equalized the intensity of light in both arms
of a differential balanced detector resulting in a zero signal. As
soon as probe polarization undergoes rotation in the sample medium,
the balance shifts towards one of the photo-diodes yielding a non-zero
signal, whose sign indicates the direction of rotation. The signal
from the amplified balanced detector was gated around the arrival
time of the probe pulse by means of a boxcar integrator, and then
sent to the lock-in amplifier.

To calibrate the absolute value of the polarization rotation angle,
we inserted the same Pockels cell in the probe beam and used it to
modulate the ellipticity of the probe polarization by a known amount
at the same frequency and under the same experimental conditions.
To convert the modulated ellipticity into the oscillations of the
polarization angle, an additional quarter-wave plate was added after
the Pockels cell (with the optical axes of the two elements oriented
at $45^{\circ}$ with respect to each other). Finally, to eliminate
possible systematic errors, each measurement was repeated with two
orthogonal states of linear probe polarization. Since the two probe
polarizations should result in signals with opposite signs, the final
drag angle was calculated as half the difference between the two signals. 
\begin{figure}
\begin{centering}
\includegraphics{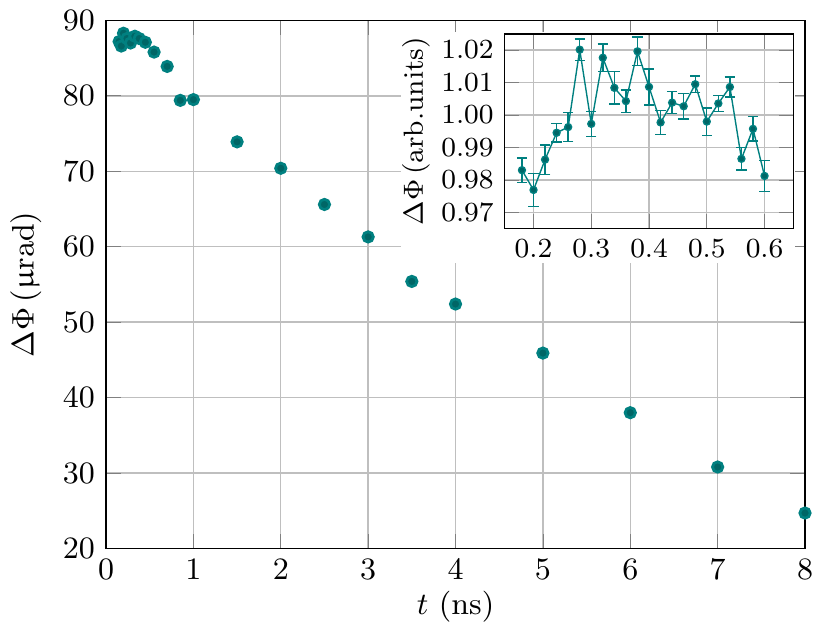}
\par\end{centering}
\caption{Experimentally measured decay of the polarization drag signal in the gas of oxygen super-rotors under the pressure of $0.5\,\mathrm{bar}$.
The centrifuge was adjusted for the terminal rotation frequency of
$7.2\,\mathrm{THz}$. The inset in the upper right corner shows the separately measured high resolution signal between $200\,\mathrm{ps}$ and $600\,\mathrm{ps}$,
where it was normalized to the average drag value in that time window.
Vertical error bars represent the standard error of the mean over
$\sim10^{5}$ laser pulses. \label{fig:FIG7}}
\end{figure}

Our experimental results are shown in Fig.~\ref{fig:FIG7}. Here,
the gas of oxygen molecules was kept under the pressure of $0.5\,\mathrm{bar}$.
By piercing the centrifuge around $75\,\mathrm{ps}$ from its front
edge, the rotational frequency of the centrifuged molecules was set
at $7.2\,\mathrm{THz}$ (as found from the corresponding Raman spectra). The observed time dependence of the polarization rotation angle reproduces quite well the results of our theoretical analysis shown in Fig.~\ref{fig:FIG3}. A transient increase of the drag signal by a few percent during the first three hundred picoseconds after the centrifuge turn off is clearly seen in the experimental data (inset in Fig. \ref{fig:FIG7}). As discussed in Section~\ref{sec:Relaxation}, this bump at the beginning of the curve is a result of a quickly decaying contribution of the slow molecular rotors, spinning in the \emph{opposite direction} with respect to the centrifuge rotation. Note that the shape of the experimentally recorded time dependence is slightly different from the theoretical predictions (both in the exact magnitude of the early bump, and the long-term decay profile) due to the nonuniform profile of the laser beam, which is hard to quantify and therefore to take into account in the numerical simulations by calculating several peak intensities (see Fig. \ref{fig:FIG3}). Moreover, the
predicted relaxation dynamics are very sensitive to both the intensity of the pulse (see Fig. \ref{fig:FIG3}) and the repulsive part of the intermolecular potential (see Appendix \ref{app:potential}), two poorly known input parameters of the simulations.
\vspace{-5mm}
\section{Conclusions \label{sec:Conclusions}}
We presented a detailed theoretical and experimental study of the recently predicted \citep{Steinitz2020} and experimentally demonstrated \citep{Milner2021} mechanical Faraday effect in a gas of fast-spinning molecules.
This phenomenon, related to Fermi's prominent polarization drag effect \citep{Fermi1923}, was examined here in an ensemble of molecules spun up by an optical centrifuge and brought to the super-rotor state that retains its rotation for a relatively long time.
To get a deeper qualitative insight into the physics of the polarization drag, we considered a simple model of molecular polarizability that accounts for both the rotational Doppler effect and Coriolis force.
This treatment was combined with molecular-dynamics simulations to account for the collisional effects in the gas. Notably, our approach not only caught the main qualitative features of the time-dependent polarization drag signals measured in our experiments, but also accurately described them quantitatively. 
In particular, our study explained a non-monotonic time-dependence of the polarization drag angle observed at the early stages of the field-free evolution, just after the end of the optical centrifuge pulses, and provided a detailed information about the long-time decay of this angle.  
As was discussed before \citep{Steinitz2012,Steinitz2016}, collisional relaxation of the unidirectional molecular rotation results in the appearance of vortex flows in gases, and the observed time-dependence of the polarization drag may be considered as an evident manifestation of this phenomenon.

Apart from the experimental demonstration of a fundamental physical effect, the presented results have several practical aspects. Polarization drag measurements may be regarded as a new diagnostic method for characterizing molecular rotation and its relaxation, an addition to the continuously expanding toolbox (e.g., \citep{Lin2015,Mizuse2015,He2019,Karamatskos2019,Bert2020} to name just a few). 
In particular, the observed non-exponential decay of the drag angle reflects the inhibited relaxation rate of the 
fast spinning molecular super-rotors and depicts their gradual thermalization over a long time-scale.
We also demonstrate how this new ability to measure rotational decay may provide valuable information on intermolecular interaction potential.
\vspace{-5mm}
\begin{acknowledgments}
\vspace{-2mm}
This work was partially supported by the
Israel Science Foundation (Grant No. 746/15)  and Canada Foundation for Innovation. I. A.
acknowledges support as the Patricia Elman Bildner
Professorial Chair and thanks the UBC Department of
Physics and Astronomy for hospitality extended to him
during his sabbatical stay.
\end{acknowledgments}
\appendix

\section{Polarizabilities - 3D case \label{sec:Polarizabilities-3D-case}}

In this Appendix, we derive the polarizability of an ensemble of polarization
charges of mass $m$ and having charge $q$. Each charge experiences
restoring forces applied by the molecular nuclei
\begin{equation}
\begin{split}\mathbf{F}_{x,\mathrm{r}} & =-k_{1}x\hat{\mathbf{x}},\;\mathbf{F}_{y,\mathrm{r}}=-k_{2}y\hat{\mathbf{y}},\;\mathbf{F}_{z,\mathrm{r}}=-k_{3}z\hat{\mathbf{z}}.\end{split}
\label{eq:App-Rest.-forces}
\end{equation}
where $\hat{\mathbf{x}}$, $\hat{\mathbf{y}}$ and $\hat{\mathbf{z}}$
are unit vectors along the three axes of a body-fixed frame associated
with each molecule. The $z$ axis coincides with the direction of
the molecular angular momentum vector $\mathbf{L}$. The molecular
bond coincides with the $x$ axis. Linear molecules rotate uniformly
with angular velocity $\Omega$ in the plane perpendicular to vector
$\mathbf{L}$. The orientation of the $z$ axis in the laboratory
fixed $XYZ$ frame is arbitrary. 

The input electric field is modeled using $\mathbf{E}=2E_{0}\cos(\omega t)\hat{\mathbf{X}}$,
or in complex notation
\begin{equation}
\mathbf{E}=E_{0}e^{i\omega t}\hat{\mathbf{X}}.\label{eq:App-electric-field}
\end{equation}

In the rotating frame, there are inertial forces acting on the mass
\begin{equation}
\mathbf{F}_{\mathrm{i}}=-2m\boldsymbol{\Omega}\times\mathbf{v}-m\boldsymbol{\Omega}\times(\boldsymbol{\Omega}\times\mathbf{r}),\label{eq:App-F-fict.}
\end{equation}
including the Coriolis (first term) and centrifugal (second term) force. Here, $\boldsymbol{\Omega}=\Omega\hat{\mathbf{z}}$, $\mathbf{r}=x\hat{\mathbf{x}}+y\hat{\mathbf{y}}+z\hat{\mathbf{z}}$,
$\mathbf{v}=\dot{\mathbf{r}}=\dot{x}\hat{\mathbf{x}}+\dot{y}\hat{\mathbf{y}}+\dot{z}\hat{\mathbf{z}}$.
Substitution yields
\begin{equation}
\mathbf{F}_{\mathrm{i}}=2m\Omega(\dot{y}\hat{\mathbf{x}}-\dot{x}\hat{\mathbf{y}})+m\Omega^{2}(x\hat{\mathbf{x}}+y\hat{\mathbf{y}}).\label{eq:App-F-fict-2}
\end{equation}

\begin{figure}[h]
\begin{centering}
\if\flag1\includegraphics{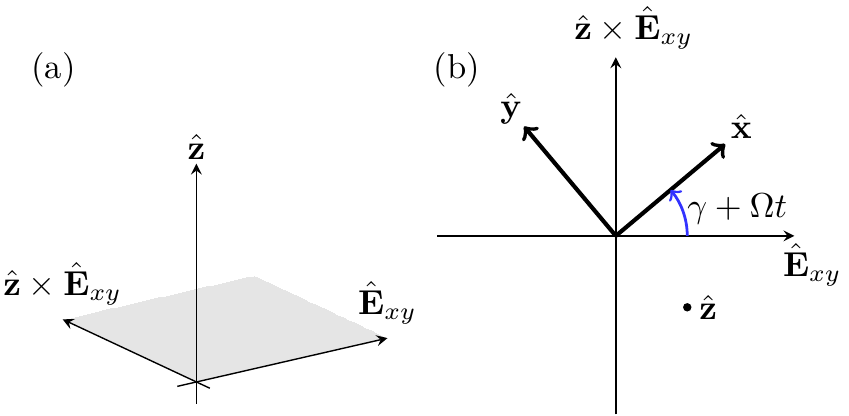}\else\include{FIG9}\fi
\par\end{centering}
\caption{(a) Basis vectors, see Eq. \eqref{eq:App-B-basis}. Notice that $\hat{\mathbf{z}}$
has an arbitrary orientation in space. (b) The $x$ and $y$ axes
rotate in the plane spanned by $\hat{\mathbf{E}}_{xy}$ and $\hat{\mathbf{z}}\times\hat{\mathbf{E}}_{xy}$.
\label{fig:FIG8}}
\end{figure}

For our purposes, it is convenient to represent $\mathbf{E}$ in a
basis $B$, which includes vector $\hat{\mathbf{z}}$, $\hat{\mathbf{E}}_{xy}$
(normalized projection of $\mathbf{E}$ on the $xy$ plane), and the
third vector formed by a cross product of the preceding two vectors
{[}see Fig. \ref{fig:FIG8}(a){]}, i.e.
\begin{equation}
B=\{\mathbf{B}_{1},\mathbf{B}_{2},\mathbf{B}_{3}\}=\{\hat{\mathbf{E}}_{xy},\hat{\mathbf{z}}\times\hat{\mathbf{E}}_{xy},\hat{\mathbf{z}}\}.\label{eq:App-B-basis}
\end{equation}

In terms of basis $B$, the electric field reads
\begin{align}
\mathbf{E} & =(\hat{\mathbf{E}}_{xy}\cdot\mathbf{E})\hat{\mathbf{E}}_{xy}+[\mathbf{E}\cdot(\hat{\mathbf{z}}\times\hat{\mathbf{E}}_{xy})](\hat{\mathbf{z}}\times\hat{\mathbf{E}}_{xy})\nonumber \\
 & +(\hat{\mathbf{z}}\cdot\mathbf{E})\hat{\mathbf{z}},\label{eq:App-E-B-repres.}
\end{align}
where
\begin{equation}
\mathbf{E}_{xy}=\mathbf{E}-(\mathbf{E}\cdot\hat{\mathbf{z}})\hat{\mathbf{z}},\quad\hat{\mathbf{E}}_{xy}=\frac{\mathbf{E}_{xy}}{|\mathbf{E}_{xy}|}\label{eq:App-Exy}
\end{equation}

\noindent and $|\cdot|$ denotes the magnitude of a vector. We want
to find the electric field projections along the rotating $x$ and
$y$ axes. As shown by Fig. \ref{fig:FIG8}(b), the $x$ axis
rotates in the plane with angular velocity $\Omega$, therefore the
relative angle between the $x$ axis and $\mathbf{E}_{xy}$ increases
with time, such that the electric field projections are
\begin{equation}
\begin{split}E_{x}= & E_{0}|[\hat{\mathbf{X}}-(\hat{\mathbf{X}}\cdot\hat{\mathbf{z}})\hat{\mathbf{z}}]|e^{i\omega t}\cos(\gamma+\Omega t),\\
E_{y}= & -E_{0}|[\hat{\mathbf{X}}-(\hat{\mathbf{X}}\cdot\hat{\mathbf{z}})\hat{\mathbf{z}}]|e^{i\omega t}\sin(\gamma+\Omega t),
\end{split}
\label{eq:App-Ex-Ey}
\end{equation}
where $\gamma+\Omega t$ is the instantaneous angle between $\hat{\mathbf{x}}$
and $\hat{\mathbf{E}}_{xy}$, and the angle $\gamma$ is defined by
\begin{equation}
\cos(\gamma+\Omega t)=\hat{\mathbf{x}}\cdot\hat{\mathbf{E}}_{xy}\label{eq:App-cos(beta)}
\end{equation}
To simplify the notation, we define two constants $c_{1}$ and $c_{2}$
\begin{equation}
\begin{split}c_{1} & \equiv|[\hat{\mathbf{X}}-(\hat{\mathbf{X}}\cdot\hat{\mathbf{z}})\hat{\mathbf{z}}]|\\
c_{2} & \equiv\hat{\mathbf{X}}\cdot\hat{\mathbf{z}}
\end{split}
\label{eq:App-constants-c}
\end{equation}
Combining the restoring forces [see Eq. \eqref{eq:App-Rest.-forces}],
fictitious forces [see Eq. \eqref{eq:App-F-fict-2}], and the force
due to the electric field, yields the following coupled equations
of motion
\begin{equation}
\begin{split}\negthickspace\negthickspace m\ddot{x}= & (m\Omega^{2}\!-\!k_{1})x\!+\!2m\Omega\dot{y}\!+\!qE_{0}c_{1}e^{i\omega t}\cos(\gamma\!+\!\Omega t),\\
\negthickspace\negthickspace m\ddot{y}= & (m\Omega^{2}\!-\!k_{2})y\!-\!2m\Omega\dot{x}\!-\!qE_{0}c_{1}e^{i\omega t}\sin(\gamma\!+\!\Omega t),\\
\negthickspace\negthickspace m\ddot{z}= & -k_{3}z+qc_{2}e^{i\omega t}.
\end{split}
\label{eq:App-eq.-of-motion-1}
\end{equation}
We solve this set of equations and then
\begin{enumerate}
\item We multiply $x(t)$, $y(t)$ and $z(t)$ by $q$ to get the induced
dipole moment components in the molecular frame
\item Express the dipole in terms of the $B$ basis
\[
q\overset{\text{\text{\tiny\ensuremath{\bm{\leftrightarrow}}}}}{\mathbf{R}}_{z}(\gamma+\Omega t)\left[\begin{array}{c}
x(t)\\
y(t)
\end{array}\right],
\]
where $\overset{\text{\text{\tiny\ensuremath{\bm{\leftrightarrow}}}}}{\mathbf{R}}_{z}(\gamma+\Omega t)$
is the canonical rotation matrix about the $z$ axis.
\item Average over the angle $\gamma$ {[}see Eq. \eqref{eq:App-cos(beta)}{]},
assuming that it is uniformly distributed in the interval $[0,2\pi]$.
\item Expand the resulting expression in a power series up to the first
order in $\Omega$.
\item Substitute $k_{1}=m\omega_{\parallel}^{2},\,k_{2}=m\omega_{\perp}^{2},\,k_{3}=m\omega_{\perp}^{2}$.
\end{enumerate}
After carrying out the listed manipulations, the induced dipole components [see Eq. \eqref{eq:App-B-basis} and
Fig. \ref{fig:FIG8}] read
\begin{equation}
\begin{split}d_{B_{1}} & =-\frac{c_{1}q^{2}E_{0}e^{i\omega t}(2\omega^{2}-\omega_{\parallel}^{2}-\omega_{\perp}^{2})}{2m(\omega_{\parallel}^{2}-\omega^{2})(\omega_{\perp}^{2}-\omega^{2})},\\
d_{B_{2}} & =\frac{ic_{1}E_{0}q^{2}e^{i\omega t}\omega(\omega_{\parallel}^{2}-\omega_{\perp}^{2}){}^{2}\Omega}{m(\omega_{\parallel}^{2}-\omega^{2}){}^{2}(\omega_{\perp}^{2}-\omega^{2}){}^{2}},\\
d_{B_{3}} & =\frac{c_{2}q^{2}E_{0}e^{i\omega t}}{m(\omega_{\perp}^{2}-\omega^{2})}.
\end{split}
\label{eq:App-dipole-projections-B}
\end{equation}
Overall, the induced dipole vector reads
\begin{equation}
\mathbf{d}=d_{B_{1}}\hat{\mathbf{E}}_{xy}+d_{B_{2}}(\hat{\mathbf{z}}\times\hat{\mathbf{E}}_{xy})+d_{B_{3}}\hat{\mathbf{z}}.\label{eq:App-induced-dipole-vector}
\end{equation}

The polarizability contribution along the $X$ axis is given by
\begin{align}
\alpha_{XX}=\frac{\hat{\mathbf{X}}\cdot\mathbf{d}}{E_{0}e^{i\omega t}} & =-\frac{q^{2}(2\omega^{2}-\omega_{\parallel}^{2}-\omega_{\perp}^{2})}{2m(\omega_{\parallel}^{2}-\omega^{2})(\omega_{\perp}^{2}-\omega^{2})}[1-(\hat{\mathbf{X}}\cdot\hat{\mathbf{z}})^{2}]\nonumber \\
 & +\frac{q^{2}}{m(\omega_{\perp}^{2}-\omega^{2})}(\hat{\mathbf{X}}\cdot\hat{\mathbf{z}})^{2},\label{eq:App-alphaXX}
\end{align}
where we used
\begin{equation}
\begin{split}\hat{\mathbf{X}}\cdot\hat{\mathbf{E}}_{xy}= & \frac{\hat{\mathbf{X}}\cdot[\hat{\mathbf{X}}-(\hat{\mathbf{X}}\cdot\hat{\mathbf{z}})\hat{\mathbf{z}}]}{|[\hat{\mathbf{X}}-(\hat{\mathbf{X}}\cdot\hat{\mathbf{z}})\hat{\mathbf{z}}]|}=\frac{1-(\hat{\mathbf{X}}\cdot\hat{\mathbf{z}})^{2}}{c_{1}},\\
\hat{\mathbf{X}}\cdot(\hat{\mathbf{z}}\times\hat{\mathbf{E}}_{xy}) & =\frac{\hat{\mathbf{X}}\cdot(\hat{\mathbf{z}}\times\hat{\mathbf{X}})}{c_{1}}=0.
\end{split}
\label{eq:App-projections-1}
\end{equation}
The polarizability contribution along the $Y$ axis reads
\begin{align}
\alpha_{XY}=\frac{\hat{\mathbf{Y}}\cdot\mathbf{d}}{E_{0}e^{i\omega t}} & =\frac{q^{2}(2\omega^{2}-\omega_{\parallel}^{2}-\omega_{\perp}^{2})}{2m(\omega_{\parallel}^{2}-\omega^{2})(\omega_{\perp}^{2}-\omega^{2})}(\hat{\mathbf{X}}\cdot\hat{\mathbf{z}})(\hat{\mathbf{Y}}\cdot\hat{\mathbf{z}})\nonumber \\
 & +\frac{iq^{2}\omega(\omega_{\parallel}^{2}-\omega_{\perp}^{2}){}^{2}\Omega}{m(\omega_{\parallel}^{2}-\omega^{2}){}^{2}(\omega_{\perp}^{2}-\omega^{2}){}^{2}}(\hat{\mathbf{Z}}\cdot\hat{\mathbf{z}})\nonumber \\
 & +\frac{q^{2}}{m(\omega_{\perp}^{2}-\omega^{2})}(\hat{\mathbf{X}}\cdot\hat{\mathbf{z}})(\hat{\mathbf{Y}}\cdot\hat{\mathbf{z}}),\label{eq:App-alphaXY}
\end{align}
where we used
\begin{equation}
\begin{split}\hat{\mathbf{Y}}\cdot\hat{\mathbf{E}}_{xy}=\frac{\hat{\mathbf{Y}}\cdot[\hat{\mathbf{X}}-(\hat{\mathbf{X}}\cdot\hat{\mathbf{z}})\hat{\mathbf{z}}]}{|[\hat{\mathbf{X}}-(\hat{\mathbf{X}}\cdot\hat{\mathbf{z}})\hat{\mathbf{z}}]|} & =-\frac{(\hat{\mathbf{X}}\cdot\hat{\mathbf{z}})(\hat{\mathbf{Y}}\cdot\hat{\mathbf{z}})}{c_{1}},\\
\negthickspace\hat{\mathbf{Y}}\cdot(\hat{\mathbf{z}}\times\hat{\mathbf{E}}_{xy})=\frac{\hat{\mathbf{Y}}\cdot(\hat{\mathbf{z}}\times\hat{\mathbf{X}})}{c_{1}} & =\frac{(\hat{\mathbf{Z}}\cdot\hat{\mathbf{z}})}{c_{1}}.
\end{split}
\label{eq:App-projections-2}
\end{equation}
We express $\alpha_{XX}$ and $\alpha_{XY}$ in terms of the angular
velocity vector ($\boldsymbol{\Omega}=\Omega\hat{\mathbf{z}}$) projections
$(\Omega_{X},\Omega_{Y},\Omega_{Z})$ and the polarizabilities $\alpha_{\parallel}(\omega)$
and $\alpha_{\perp}(\omega)$ of a non-rotating molecule
\begin{equation}
\alpha_{\parallel}(\omega)=\frac{q^{2}}{m(\omega_{\parallel}^{2}-\omega^{2})},\quad\alpha_{\perp}(\omega)=\frac{q^{2}}{m(\omega_{\perp}^{2}-\omega^{2})}.\label{eq:App-stationary-polarizabilities}
\end{equation}
Then, we carry out averaging over the angular velocities
\begin{align}
\braket{\alpha_{XX}} & =\frac{1}{2}[\alpha_{\parallel}(\omega)+\alpha_{\perp}(\omega)]\nonumber \\
 & -\frac{1}{2}[\alpha_{\parallel}(\omega)-\alpha_{\perp}(\omega)]\Braket{\frac{\Omega_{X}^{2}}{\Omega^{2}}},\label{eq:App-alphaXX-1}
\end{align}
\begin{align}
\braket{\alpha_{XY}} & =-\frac{1}{2}(\alpha_{\parallel}-\alpha_{\perp})\Braket{\frac{\Omega_{X}\Omega_{Y}}{\Omega^{2}}}\nonumber \\
 & +\frac{i}{2}(\alpha_{\parallel}^{\prime}-\alpha_{\perp}^{\prime})\frac{\alpha_{\parallel}-\alpha_{\perp}}{\alpha_{\parallel}+\alpha_{\perp}}\braket{\Omega_{Z}}.\label{eq:App-alphaXY-1}
\end{align}
Similarly, it can be shown that
\begin{align}
\braket{\alpha_{YY}} & =\frac{1}{2}[\alpha_{\parallel}(\omega)+\alpha_{\perp}(\omega)],\nonumber \\
 & -\frac{1}{2}[\alpha_{\parallel}(\omega)-\alpha_{\perp}(\omega)]\Braket{\frac{\Omega_{Y}^{2}}{\Omega^{2}}},\label{eq:App-alphaYY}
\end{align}
and $\braket{\alpha_{XY}}=\braket{\alpha_{XY}^{\dagger}}$. Notice
that in the case of isotropic molecular ensemble ($\braket{\Omega_{X}^{2}/\Omega^{2}}=\braket{\Omega_{Y}^{2}/\Omega^{2}}=1/3$
and $\braket{\Omega_{X}\Omega_{Y}/\Omega^{2}}=0$), we get the well
known result, $\braket{\alpha_{XX}}=\braket{\alpha_{YY}}=(\alpha_{\parallel}+2\alpha_{\perp})/3$,
while the off diagonal elements vanish.

Assuming $\braket{\Omega_{X}\Omega_{Y}/\Omega^{2}}\approx0$, Eq.
\eqref{eq:App-alphaXY-1} reduces to
\begin{equation}
\braket{\alpha_{XY}}=\frac{i}{2}(\alpha_{\parallel}^{\prime}-\alpha_{\perp}^{\prime})\frac{\alpha_{\parallel}-\alpha_{\perp}}{\alpha_{\parallel}+\alpha_{\perp}}\braket{\Omega_{Z}}.\label{eq:App-<alphaXY>}
\end{equation}
Finally, substitution into Eq. \eqref{eq:polarization-rotation} yields
\begin{equation}
\Delta\Phi=\frac{1}{2}(\alpha_{\parallel}^{\prime}-\alpha_{\perp}^{\prime})\frac{\alpha_{\parallel}-\alpha_{\perp}}{\alpha_{\parallel}+\alpha_{\perp}}\frac{\omega NL}{2c\varepsilon_{0}}\braket{\Omega_{Z}}.\label{eq:App-Delta-Phi-1}
\end{equation}
When $\alpha_{\parallel}=\alpha_{\perp}$, i.e. when the tensor of
polarizability is isotropic, $\Delta\Phi=0$. This is consistent with
the results of \citep{Baranova1979}. When $\alpha_{\perp}=0$, i.e.
the charge is restricted to move along the molecular axis,
\begin{equation}
\Delta\Phi=\frac{\alpha_{\parallel}^{\prime}}{2}\frac{\omega NL}{2c\varepsilon_{0}}\braket{\Omega_{Z}},\label{eq:App-=00005CDelta=00005CPhi-alphas-2}
\end{equation}
which is the same as in \citep{Steinitz2020}.

\section{Frequency dependent polarizabilities of oxygen molecule \label{sec:Appendix-Frequency-dep-pol}}

The evaluation of the angle of polarization rotation {[}see Eq. \eqref{eq:App-Delta-Phi-1}{]}
requires $\alpha_{\parallel}(\omega)$ and $\alpha_{\perp}(\omega)$,
where $\omega$ is the frequency of the driving (probe) field, and
their derivatives. According to \citep{Hettema1994} (see Eq. 4, Table
IV), the frequency-dependent polarizabilities (in atomic units) of
molecular oxygen ($\mathrm{O}_2$) are given by
\begin{align}
\alpha_{\parallel}(\omega) & =\sum_{k=0}^{3}S_{\parallel}^{(-2k-2)}\omega{}^{2k},\label{eq:pol-parallel}\\
\alpha_{\perp}(\omega) & =\sum_{k=0}^{3}S_{\perp}^{(-2k-2)}\omega{}^{2k}.\label{eq:pol-perpendicular}
\end{align}
Note that the superscripts $(-2k-2)$ are indices, while $\omega{}^{2k}$
actually means ``omega raised to the power $2k$''. The numerical
values of the coefficients (in atomic units) $S_{\parallel,\perp}^{(-2)},\,S_{\parallel,\perp}^{(-4)},\,S_{\parallel,\perp}^{(-6)},\:S_{\parallel,\perp}^{(-8)}$
are
\begin{align}
S_{\parallel} & =[14.993,\,67.040,\,520.995,\,4680.275],\label{eq:Spar}\\
S_{\perp} & =[7.834,\:13.988,\,39.826,\,151.640],\label{eq:Sper}
\end{align}
for $k=0,\dots,3$ (see Table IV in \citep{Hettema1994}). Higher
order coefficients, $S_{\parallel,\perp}^{(-10)},S_{\parallel,\perp}^{(-12)}\dots$,
are neglected. For $\omega=0$, the polarizabilities are $\alpha_{\parallel}(0)=S_{\parallel}^{(-2)}\approx15\,\mathrm{a.u.}$
and $\alpha_{\perp}(0)=S_{\perp}^{(-2)}\approx7.8\,\mathrm{a.u.}$,
which is close to the values for $\mathrm{O}_2$ molecule provided in NIST database \citep{NIST} (calculated
polarizabilities). In our experiments, we used a $\sim400\,\mathrm{nm}$ wavelength
probe pulse, which corresponds to $\omega=0.11\,\mathrm{a.u.}$ The
approximate values of polarizabilities and their derivatives at $\omega=0.11\,\mathrm{a.u.}$
are $\alpha{}_{\parallel}=15.80\,\mathrm{a.u.}$, $\alpha{}_{\parallel}=8.00\,\mathrm{a.u.}$,
$\alpha{}_{\parallel}'=18.04\,\mathrm{a.u.}$, $\alpha{}_{\perp}'=3.30\,\mathrm{a.u.}$

\section{Dependence on the intermolecular interaction potential} \label{app:potential}
To test the sensitivity of the calculated time dependence of the drag angle on the intermolecular forces, we carried calculations
using three different $\mathrm{O}_{2}$--$\mathrm{O}_{2}$ atom-atom
anisotropic potentials, modeled using
\begin{align}
\negthickspace\negthickspace V_{n-6} & =\sum_{i=1,2}\sum_{j=1,2}4\varepsilon_{n-6}\left[\left(\frac{\sigma_{n-6}}{R_{ij}}\right)^{n}-\left(\frac{\sigma_{n-6}}{R_{ij}}\right)^{6}\right]\nonumber \\
 & +\frac{q_{i}q_{j}}{R_{ij}},\label{eq:Lennard-Jones}
\end{align}
where $i$ and $j$ run over the atoms of the first and second molecule,
respectively. Here $n=12$ (the usual 12-6 Lennard-Jones potential),
$10$, or $14$. $R_{ij}$ is the distance between atom $i$ of the
first molecule and atom $j$ of the second molecule. The ``reference''
potential used, which corresponds to the choice $n=12$, was proposed
in \citep{Bouanich1992} where the $\mathrm{O}_{2}$ electric quadrupole
(i.e. the charges $q_i$) was fixed and the $\varepsilon_{12-6}$ and $\sigma_{12-6}$
parameters for O--O interactions were obtained from fits of measured
second-virial coefficients at temperatures between $200\,\mathrm{K}$
and $400\,\mathrm{K}$. This potential was later on used in requantized
Molecular Dynamics Simulations of the shape of pure $\mathrm{O}_{2}$
absorption lines, leading to excellent agreements with measured data
\citep{Hartmann2013b}. Starting from this $V_{12-6}$ potential,
we generated $V_{10-6}$ and $V_{14-6}$ potentials, with $\sigma_{12-6}=\sigma_{10-6}=\sigma_{14-6}$
and associated values of $\varepsilon_{10-6}$ and $\varepsilon_{14-6}$
for which the O--O interaction has the same minimum value $-\varepsilon_{12-6}$
(note that since the same $\sigma$ is used, the three atom-atom potentials
are also identical and equal to zero for the distance $R_{ij}=\sigma_{12-6}$).

\begin{figure}[h]
\begin{centering}
\includegraphics{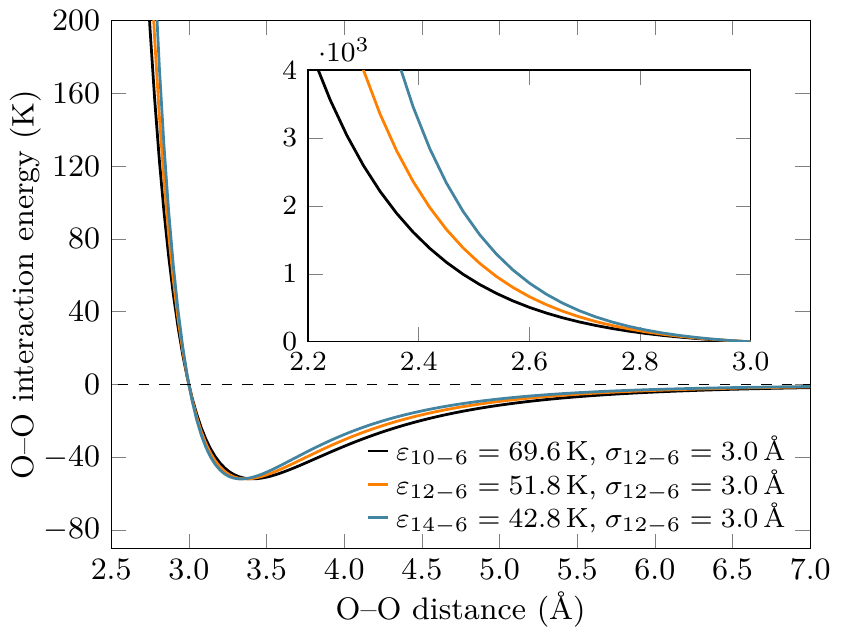}
\par\end{centering}
\caption{Interatomic interaction potential between two oxygen atoms, see Eq.
\eqref{eq:Lennard-Jones}. \label{FIG9}}
\end{figure}

The three potentials are plotted in Fig. \ref{FIG9},
which shows that in the range where the potential is comparable to
the kinetic temperature, the values are very close. This implies that
using these three potentials would lead, for quantities essentially
sensitive to the potential well and lower part of the repulsive front
(such as the virial coefficients and line shapes near room temperature
mentioned above, and the isotropization of the angular momentum of
noncentrifuged molecules as discussed below), to very similar results.

In contrast, as expected, the inset in Fig. \ref{FIG9}
shows that in the repulsive region with large interaction energies,
differences between the three potentials are apparent. The 10-6 (resp.
14-6) potential being significantly weaker and less steep (resp. larger
and steeper) than the 12-6 potential. Intermolecular forces being
given by the gradient of the potential, it is obvious that they significantly
increase (for collisions involving short distances) when going from
the 10-6 to the 14-6 potential, with the 12-6 potential being in between
these two. One can thus expect that the three potentials should lead
to significantly different decay rates (including
that of the angular momentum) of molecules which have been centrifuged
up to very high angular velocities (magnitude and orientation)
since changing their angular velocities requires extremely efficient
collisions which only occur when the strongly repulsive part of the
potential is involved (i.e. short intermolecular distances). 

\begin{figure}
\begin{centering}
\vspace{3mm}
\includegraphics{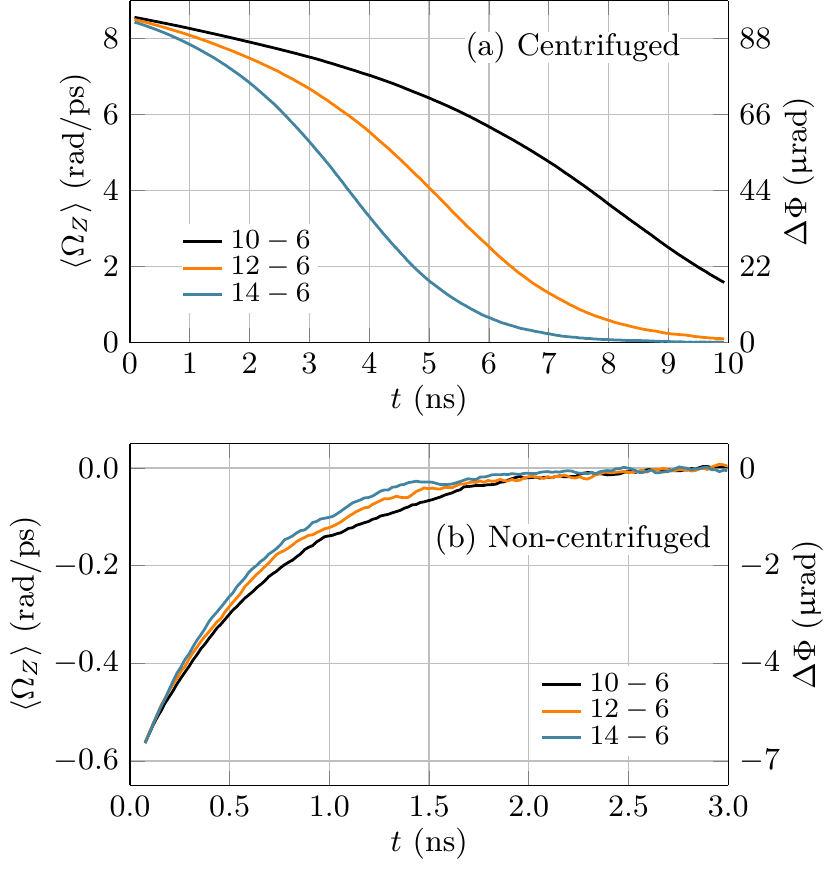}
\par\end{centering}
\caption{Average $Z$ projection of the molecular angular velocity (left ordinate)
and angle of polarization rotation (right ordinate) for a $\tau=75\,\mathrm{ps}$
long centrifuge having peak intensity of $I_{0\mathrm{c}}=3.5\,\mathrm{TW/cm^{2}}$
applied to pure $\mathrm{O}_{2}$ gas at temperature $T=300\,\mathrm{K}$
and pressure $0.5\,\mathrm{bar}$. The propagation length is $L=1\,\mathrm{mm}$.
(a) Centrifuged molecules (b) Non-centrifuged molecules. \label{FIG10}}
\end{figure}

Obviously,
in this case, the dissipation time should decrease when going from
the 10-6 to the 14-6 potential. The above qualitative arguments are confirmed by Fig. \ref{FIG10}. The latter
displays, for $\mathrm{O}_{2}$ gas at $0.5\,\mathrm{bar}$, the average
$Z$ components of the angular velocity after excitation by a $75\,\mathrm{ps}$
long optical centrifuge with peak intensity of $3.5\,\mathrm{TW/cm^{2}}$. Similar to Fig. \ref{fig:FIG5}, the contributions of centrifuged
and non-centrifuged molecules have been separated. As expected, the various potentials lead to very close results in the latter case, while significant differences are obtained in the former. These results suggest that studying the thermalization of optically-centrifuged molecules appears as an interesting tool to test (and improve) the strongly
repulsive region of intermolecular potentials.

\end{document}